\documentclass[conference]{IEEEtran}
\IEEEoverridecommandlockouts
\def\abbrtitle{LATTE}
\def\fulltitle{Low-Precision Approximate Attention with Head-wise Trainable Threshold for Efficient Transformer}
\def\boldtitle{\textbf{L}ow-Precision \textbf{A}pproximate Attention with Head-wise \textbf{T}rainable \textbf{T}hreshold for \textbf{E}fficient Transformer}
\def\ultitle{\underline{L}ow-Precision \underline{A}pproximate Attention with Head-wise \underline{T}rainable \underline{T}hreshold for \underline{E}fficient Transformer}
\usepackage{cite}
\usepackage{amsmath,amssymb,amsfonts}
\usepackage{graphicx}
\usepackage{textcomp}
\usepackage{xcolor}
\usepackage{subcaption}
\usepackage{hyperref}
\usepackage{hhline}
\usepackage{algorithm2e}
\RestyleAlgo{ruled}
\SetKwInOut{KwIn}{Input}
\SetKwInOut{KwOut}{Output}
\SetKwComment{Comment}{/* }{ */}
\def\BibTeX{{\rm B\kern-.05em{\sc i\kern-.025em b}\kern-.08em
    T\kern-.1667em\lower.7ex\hbox{E}\kern-.125emX}}
\renewcommand{\baselinestretch}{0.97}
\begin{document}

\title{\huge\textbf{\abbrtitle}: \boldtitle}
\author {Jiing-Ping Wang, Ming-Guang Lin and An-Yeu (Andy) Wu\\
Graduate Institute of Electronics Engineering, National Taiwan University, Taipei, Taiwan\\
\{benwang, chrislin\}@access.ee.ntu.edu.tw, andywu@ntu.edu.tw
}
\maketitle

\begin{abstract}
With the rise of Transformer models in NLP and CV domain, Multi-Head Attention has been proven to be a game-changer. However, its expensive computation poses challenges to the model throughput and efficiency, especially for the long sequence tasks. Exploiting the sparsity in attention has been proven to be an effective way to reduce computation. Nevertheless, prior works do not consider the various distributions among different heads and lack a systematic method to determine the threshold. To address these challenges, we propose \ultitle\ (\abbrtitle). \abbrtitle\ employs a head-wise threshold-based filter with the low-precision dot product and computation reuse mechanism to reduce the computation of MHA. Moreover, the trainable threshold is introduced to provide a systematic method for adjusting the thresholds and enable end-to-end optimization. Experimental results indicate \abbrtitle\ can smoothly adapt to both NLP and CV tasks, offering significant computation savings with only a minor compromise in performance. Also, the trainable threshold is shown to be essential for the leverage between the performance and the computation. As a result, \abbrtitle\ filters up to 85.16\% keys with only a 0.87\% accuracy drop in the CV task and 89.91\% keys with a 0.86 perplexity increase in the NLP task.
\end{abstract}


\section{Introduction}
In recent years, with the advantages of Multi-Head Attention, Transformer model\cite{vaswani2017attention} has illuminated NLP field, establishing itself as the state-of-the-art in a range of tasks. Models like BERT\cite{devlin2018bert}, XLNet\cite{yang2019xlnet}, and GPT\cite{radford2019language, brown2020language} series have excelled in areas including question-answering, information retrieval, machine translation, etc. Moreover, Transformer model has shown significant promise in the realm of computer vision, with ViT\cite{dosovitskiy2020image} attaining impressive accuracy in image classification tasks.

\par
However, a well-known concern of Attention is its quadratic time and memory complexity to the input sequence length. 
% The dot product of the query and key vectors inherently leads to a $\mathcal{O}(n^2)$ computational complexity, which grows rapidly as the sequence lengthens. Moreover, multiplying the attention probabilities with the value vectors also significantly adds to the computation load. Therefore, it's imperative to design an algorithm addressing the growing computation.
Several works have focused on the sparsity of Attention --- not every key holds equal importance to a query. Hence, many keys can be omitted without degrading the performance. Some studies enhance the top-\textit{k} algorithm: $A^3$\cite{ham20203} sorts the key vectors element-by-element to find the larger dot product without completing all the computations. While considering the difficulty of implementing a specialized sorting engine for the top-\textit{k} algorithm, several works move to the threshold-based approach: ELSA\cite{ham2021elsa} selects the appropriate keys based on their high similarity score to the query. Energon\cite{zhou2022energon} filters the keys by approximation with an adjustable threshold. Although the threshold-based algorithm yields promising performance, there still remain some challenges: 1) Without a systematic method to leverage between the pruning ratio and the model performance, it is time-consuming the search for the desired threshold configuration.  2) The values of Attention score vary between different blocks and heads. Therefore, a global threshold is not suitable for the whole model.

% In our study, we start with Energon, which saves the computation by employing the low-precision dot product to prune the keys. Additionally, to minimize the filter overhead, Energon utilizes a two-stage filter and adopts a 2-bit dot product as a preliminary filter followed by a 4-bit dot product exclusively on the unpruned keys as a secondary filter, so most of the computation is conducted with the lowest precision. Still, during our experiment, we found there exists some room to be improved in Energon: First, a single-round filter, when paired with an alternative filter method, is sufficient to achieve a high pruning ratio with minimal performance degradation, and it also reduces the filter overhead. Second, Energon lacks a systematic method to tune the hyperparameter. Within its algorithm, hyperparameters are adjusted by an exhaustive search, which makes it impractical to find an optimal configuration. Also, this exhaustive search does not provide any control over the desired pruning ratio.
\begin{figure}[t]
    \centering
    \includegraphics[width=0.95\linewidth]{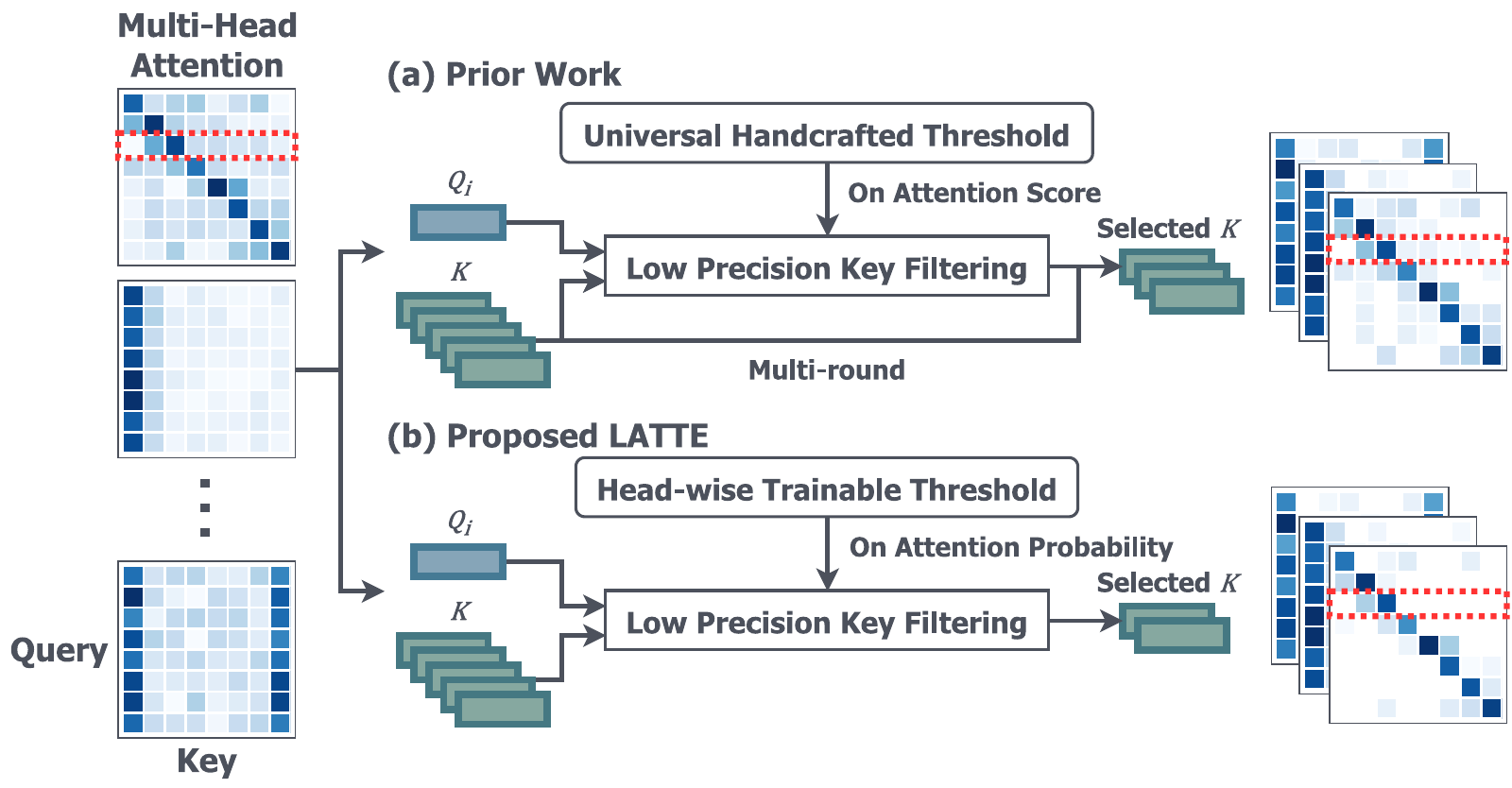}
    \caption{Comparison between (a) prior work and (b) the proposed \abbrtitle}
    \label{fig: algorithm_comp}
\end{figure}

\par
To mitigate the issues mentioned above, we proposed \fulltitle\ (\abbrtitle), a low-precision filter algorithm designed to exploit the sparsity of Attention mechanism with the trainable threshold empowering more granular control on thresholds. As shown in Fig. \ref{fig: algorithm_comp}, in our work, there are two main differences between \abbrtitle\ and prior works: 1) The thresholds in \abbrtitle\ are determined by a proportion of the post-softmax value to achieve an even better performance computation trade-off. 2) \abbrtitle\ adopts the trainable threshold to provide a systematic approach to search for the optimal configuration. With the proposed method, we can reduce 85.16\% keys with only a 0.87\% accuracy drop in the CV task and reduce 89.91\% keys with a 0.86 perplexity increase in the NLP task.
% Unlike Energon, which determines its threshold by interpolating between the minimum, maximum, and mean values, our algorithm sets the threshold by subtracting a trained parameter from the maximum values. Also, since the threshold is trained with backpropagation, we can provide more granular control on it, e.g. the desired pruning ratio being manually assigned, and we can replace the universal threshold with the head-wise thresholds. With this differential threshold, our work pruned more than 80\% keys with less than 1\% accuracy drop, using only a single-round filter (Section IV-A).

\begin{figure*}[t]
    \centering
    \includegraphics[width=0.7\linewidth]{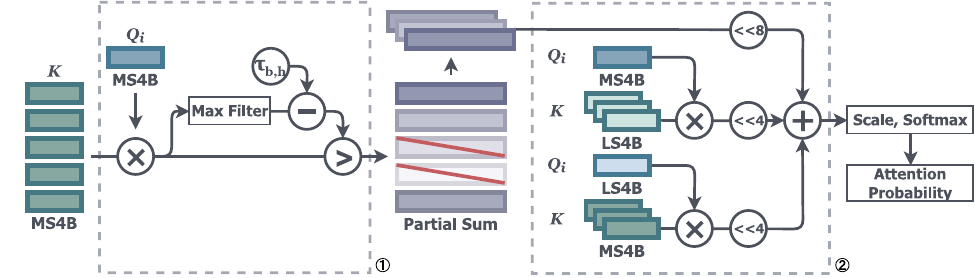}
    \caption{Overview of \abbrtitle\ algorithm. \raisebox{.5pt}{\textcircled{\raisebox{-.9pt} {1}}} The key vectors are filtered based on the low-precision dot product, and \raisebox{.5pt}{\textcircled{\raisebox{-.9pt} {2}}} the partial sum of retained keys is reused for the reduced dot product. \vspace{-1em}}
    \label{fig: algorithm_flow}
\end{figure*}

\section{Background and Related Works}
\subsection{Transformer Models}
First proposed by Google\cite{vaswani2017attention}, Transformer model has shown great potential in various NLP tasks. Many of its descendant variants become state-of-the-art as well, making it a well-known neural network architecture. Besides its excellence in NLP domain, Transformer model also outperformed traditional CNNs in many representative computer vision tasks.
\par
One reason that sets Transformer apart is its Attention mechanism. Attention comprises three linear projection weights $\{W^Q, W^K, W^V\}\in\mathcal{R}^{d_{in}\times d_k}$, where $d_{in}$ is the input dimension and $d_k$ is the embedding dimension. The input sequence denoted as $X\in\mathcal{R}^{n\times d_{in}}$, where $n$ is the sequence length, is first projected into Query, Key, and Value, $\{Q, K, V\}\in\mathcal{R}^{n\times d_k}$, by the projection weights. $Q, K, V$ are then split into $h$ heads, $Q_h,K_h,V_h$, respectively. For each head, $Q_h, K_h$ are multiplied and scaled to compute the attention score,
\begin{equation}
\footnotesize
    A_h=\frac{Q_h\cdot K_h^T}{\sqrt{d_h}}.
\end{equation}
The attention score is further passed through a $Softmax$ function to derive the attention probability, $P_h$. Multiplying this attention probability with $V_h$ yields the output of Attention.
\begin{equation}
\footnotesize
    P_h=Softmax(A_h),
\end{equation}
\begin{equation}
\footnotesize
    Attention(Q_h,K_h,V_h)=P_h\cdot V_h.
\end{equation}
\par
The pair-wise dot product in Attention makes its computational complexity $O(n^2d)$, and the matrix multiplication between $P$ and $V$ also leads to $O(n^2d)$ of complexity. In recent models, $d$ is commonly set to 768, and $n$ would be 197 or 577 for CV tasks and 1024 for NLP tasks. Thus, reducing $n$ can effectively conserve the computation in both operations.
\subsection{Related Works of Sparse Attention with Pruning}
In the computation of Attention, the dot product between the query and the key can be regarded as the similarity between the query and the key. Therefore, the attention score implies the importance of the query, key pair. By omitting the query, key pairs of lower importance, the computation can be reduced without degrading the performance of the model.
\par
The author of $A^3$\cite{ham20203} proposed top-\textit{k} styled algorithm. For each query, $A^3$ computes the dot product with the key matrix element-by-element. During each iteration, only the elements that produce the largest and smallest values are considered and aggregated. After several iterations, the approximate attention score matrix is derived. The obvious issue of $A^3$ is the requirement and the overhead of a specialized sorting engine.
\par
To avoid the inevitable sorting in the top-\textit{k} selection, other works move to the threshold-based approach. ELSA\cite{ham2021elsa} uses sign random projection (SRP, a type of locality sensitive hash) to encode each query and key to a bit stream. With the hamming distance between each encoded query and key, ELSA can approximate the similarity score by some calibration.
%:
% \begin{equation}
%     cos(\theta_{q,k})=cos(\frac{\pi}{k}hamming(q_e,k_e)-\theta_{bias}),
% \end{equation}
% where $q, k$ are a pair of query and key vector, $k$ is the length of the bit stream, $q_e, k_e$ are the encoded query and key, and $\theta_{bias}$ is calibration factor.
With the estimated attention score, ELSA then selects the keys with a predefined layer-wise threshold and computes the Attention with only these keys. Nevertheless, ELSA determines its thresholds by human-defined hyperparameters, which are based on a heuristic guess and might need several trials to find the optimal configuration.
\par
Researchers proposed Mix-Precision Multi-Round Filter (MP-MRF) approach in Energon\cite{zhou2022energon}. The scaled dot product is first computed at lower precision as an estimate of the attention score. Based on the estimation, Energon filters the keys and only retains those likely to yield higher attention scores. Additionally, to minimize the filter overhead, MP-MRF adopts a two-staged approach with a 2-bit precision dot product estimation as a preliminary filter followed by a 4-bit estimation, making most of the dot product be computed only at 2-bit precision. To avoid the inevitable sorting in the top-\textit{k} selection, Energon presents a threshold-based filter. By interpolating among the maximum, minimum, and mean values of a row of attention scores with a handcrafted universal hyperparameter $\alpha$, Energon determines a threshold for key filtering,
\begin{equation}
\footnotesize
    \theta^r_i=\begin{cases}
        \alpha_r\times max(A_i)+(1-\alpha_r)\times mean(A_i), 0\leq\alpha_r<1 \\
        -\alpha_r\times min(A_i)+(1+\alpha_r)\times mean(A_i), -1<\alpha_r<0
    \end{cases}.
\end{equation}
$\theta^r_i$ denotes the threshold for the \textit{i}-th row and r-th round, and $A_i$ denotes the attention score of the \textit{i}-th row.
Only the keys that produce attention scores higher than this threshold are retained. However, like ELSA, the performance of Energon is limited by the quality of the universal hyperparameter.

\section{Proposed Algorithm}
In this section, we proposed \fulltitle\ (\abbrtitle). The overview of \abbrtitle\ algorithm is shown in Fig. \ref{fig: algorithm_flow}, which includes low-precision approximate Attention with trainable threshold $\tau$ and computation reuse with the computed low-precision attention score. Subsection III-A depicts the low-precision approximated Attention, and subsection III-B shows a systematic hyperparameter adjustment algorithm with the trainable threshold. The overall \abbrtitle\ flow is described in Algorithm \ref{alg: LAT}.

% In this work, we focus on the following two challenges: (1) A single-round algorithm that can successfully reduce the overhead. (2) A systematic algorithm to search the hyperparameters, offering granular control over the thresholds. Therefore, we proposed the Low-Precision Filter with Trainable Threshold (LoFT). Details are discussed as follows, and the overall LoFT flow is described in Algorithm \ref{alg: LoFT}.

\subsection{Low-Precision Approximated Attention}
\paragraph{\textbf{Differential Threshold}}
Based on the observation that a few keys dominate the attention probability, most of the computation can be saved without diminishing the performance by only keeping the entries that are larger than a proportion $\gamma$ of the largest element.

% Instead of selecting the k-largest elements in the attention score, we focus on the attention probabilities after $Softmax$. In our proposed method, we only keep the elements that are larger than a proportion $\gamma$ of the largest element:
\begin{equation}
\footnotesize
    \theta_{prob,i} = \gamma max(P_i),
\end{equation}
where $\theta_{prob, i}$ denotes the threshold of the \textit{i}-th row, and $P_i$ denotes the attention probability of the \textit{i}-th row.
\par
To avoid the redundant $Softmax$ calculation on the reduced elements, the filter operation can be performed before $Softmax$, and the threshold becomes

\begin{equation}
\footnotesize
    \theta_{prob,i} = \frac{\gamma exp(max(A_i))}{\Sigma^n_{j=1}exp(A_i[j])}     =\frac{exp(max(A_i)-(-ln(\gamma)))}{\Sigma^n_{j=1}exp(A_i[j])}, \\
\end{equation}
\begin{equation}
\footnotesize
    \theta_{score,i} = max(A_i) - \tau.
\end{equation}
$A_i$ and $\theta_{score, i}$ denote the attention score and the threshold for the attention score of the \textit{i}-th row, respectively. As shown above, the proportional threshold parameter of $\gamma$ on attention probability is equivalent to the differential threshold parameter of $\tau$ on attention score. To clarify, $\theta_{prob,i}$ and $\theta_{score,i}$ are two equivalent thresholds performed on attention probability and attention score respectively, while $\gamma$ and $\tau$ are two parameters to determine the two thresholds from attention probability and attention score.

\begin{algorithm}[t]
\footnotesize
    \caption{Low-Precision Approximate Attention}
    \label{alg: LAT}
    \KwIn{Trained threshold $\tau$ \\
    8-bit quantized $Q$, $K$, $V$ $\in R^{n\times d_k}$}
    \KwOut{Approximate Attention}
    Let $A[n][n]$ be a new array\;
    $A_{estimate} \gets Q[7:4]\cdot K[7:4]^T$\;
    \For{each row i} {
        $\theta_i \gets max(A_{estimate}[i]) - \tau$\;
        \For{each column j} {
            \eIf{$A_{estimate}[i][j] > \theta_i$} {
                Reuse $A_{estimate}$ for approximate 8-bit dot product\;
                $A[i][j] \gets$ approximate 8-bit dot product\;
            } {
                $A[i][j] \gets 0$\;
            }
        }
        Compute approximate Attention with non-zero entries\;
        $P \gets Softmax(A / \sqrt{d_h})$\;
        $O \gets P\cdot V$\;
    }
    \Return $O$\;
\end{algorithm}

\paragraph{\textbf{Low-Precision Estimation}}
Our proposed approach determines the thresholds based on the maximum attention scores, which inevitably requires computing all the dot products. To reduce the overhead on the omitted elements, we estimate the attention scores with lower precision rather than computing the dot product directly. In our experiment, with the query and key quantized to 8-bit, we found that estimating the dot product with the most significant 4 bits (MS4B, and LS4B denotes the least significant 4 bits) gives a promising result as shown later.

\paragraph{\textbf{Computation Reuse}}
To further minimize the dot product overhead, we repurpose the result of the dot product with MS4B to assist in computing the 8-bit dot product. Note that an 8-bit dot product can be factorized into four 4-bit dot products,

\begin{figure}[t]
    \centering
    \begin{subfigure}[b]{0.32\linewidth}
        \centering
        \includegraphics[width=\textwidth]{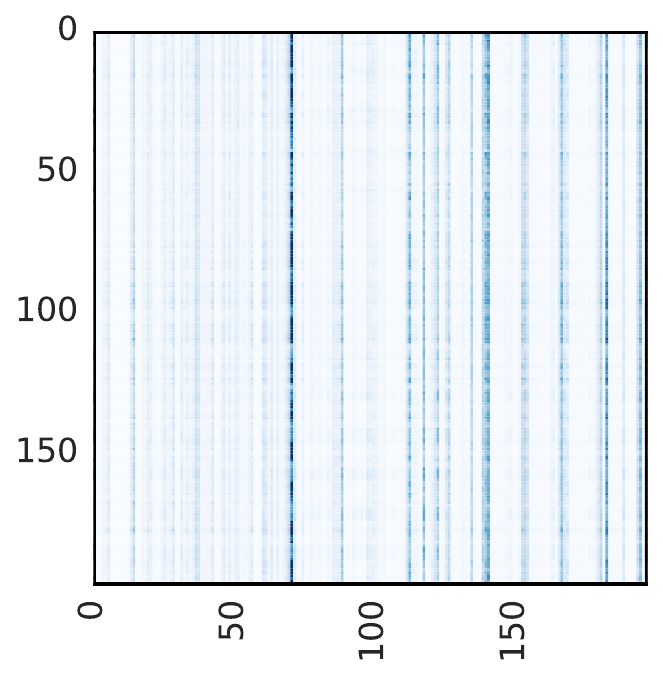}
        \caption{Block 2, Head 2}
    \end{subfigure}
    \hfill
    \begin{subfigure}[b]{0.32\linewidth}
        \centering
        \includegraphics[width=\textwidth]{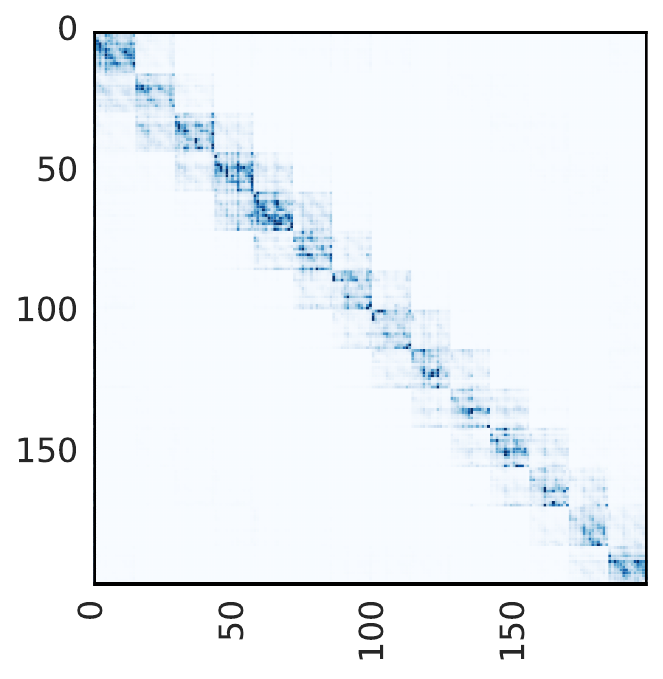}
        \caption{Block 4, Head 11}
    \end{subfigure}
    \hfill
    \begin{subfigure}[b]{0.32\linewidth}
        \centering
        \includegraphics[width=\textwidth]{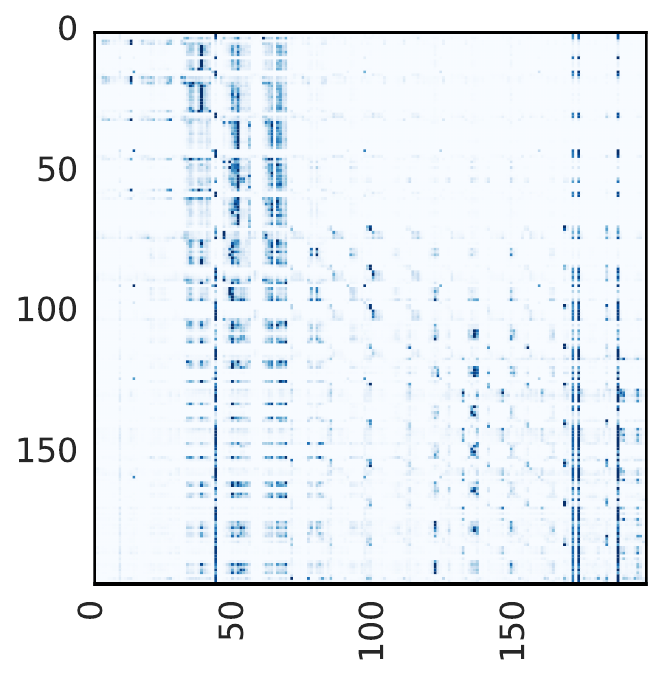}
        \caption{Block 10, Head 9}
    \end{subfigure}
    \caption{Attention probabilities among different blocks and heads of DeiT\cite{touvron2021training}}
    \label{fig: attention_probs}
\end{figure}

\footnotesize
\begin{align}
    (Q_M&\cdot2^4+Q_L)\cdot(K_M\cdot2^4+K_L)^T \\ 
    &= Q_M\cdot K_M^T\cdot2^8 + (Q_M\cdot K_L^T+Q_L\cdot K_M^T)\cdot2^4 + Q_L\cdot K_L^T,
\end{align}
\normalsize
where $Q_M$, $L_M$ is the most significant 4 bits of $Q,K$, $Q_L$, $K_L$ is the least significant 4 bits of $Q,K$, and both with range of $[-8, 7]$. The first term $Q_M\cdot K_M^T$ is exactly the estimated attention score and can be reused to reduce the computation of the dot product. In our experiment, we observed that the last term $Q_L\cdot K_L^T$ has a negligible impact on the model performance. Thus, in our algorithm, the low-precision dot product is compensated by adding $(Q_M\cdot K_L^T+Q_L\cdot K_M^T)\cdot2^4$ and serves as the approximate 8-bit dot product.

\subsection{Head-wise Trainable Threshold}
A major shortcoming of the differential threshold is the unbounded hyperparameters $\tau$. The only constraint $\tau \geq 0$ makes it impractical for an exhaustive search. Hence, there arises the necessity for a systematic approach to adjust the hyperparameters. In our proposed method, we address this issue by the trainable threshold. We freeze the parameters of the backbone model and only train the thresholds with the loss function,
\begin{equation}
\footnotesize
    \mathcal{L}oss=\alpha\mathcal{L}_{pred}+\beta\mathcal{L}_{prune}+\kappa\mathcal{L}_{KD}.
\end{equation}
$\mathcal{L}_{pred}$ denotes the distinct loss function for each task, e.g. cross-entropy for image classification in CV and log-likelihood for text generation task in NLP, $\mathcal{L}_{prune}$ is the MSE between the average pruning ratio and the target pruning ratio, and $\mathcal{L}_{KD}$ stands for the knowledge distillation loss. Using the trainable threshold, we can enhance the performance, which will be discussed later.
\par
Furthermore, as shown in Fig. \ref{fig: attention_probs}, the magnitude and distribution of the attention probability vary from head to head, and we think a universal threshold might not suffice for optimal performance. The trainable threshold also enables head-wise threshold ($\tau_{b,h}$, where $b$ denotes the block index, and $h$ denotes the head index) adjustments.

\section{Experimental Results}
To evaluate our algorithm, we choose two representative tasks from both NLP domain and CV domain as benchmarks: for the NLP task, we select GPT-2\cite{radford2019language} on WikiText-2\cite{merity2016pointer}, and for the CV task, we choose DeiT\cite{touvron2021training} on Imagenet1K\cite{deng2009imagenet}. Both the implementations are from Hugging Face\cite{Wolf_Transformers_State-of-the-Art_Natural_2020}. The trainable thresholds are trained on a calibration set of size 10k, which is randomly drawn from the training set. The target pruning ratios are chosen in $[0.4, 0.9]$ with the step of 0.1.

\subsection{Comparison with Prior Work}
\begin{figure}[t]
    \centering
    \hfill
    \begin{subfigure}[b]{0.47\linewidth}
        \centering
        \includegraphics[width=\textwidth]{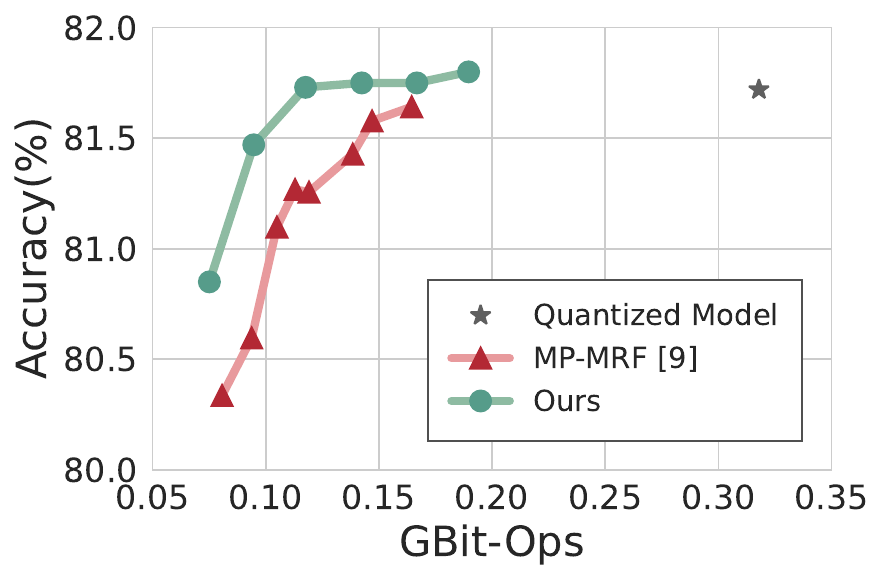}
        \caption{CV task}
    \end{subfigure}
    \hfill
    \begin{subfigure}[b]{0.47\linewidth}
        \centering
        \includegraphics[width=\textwidth]{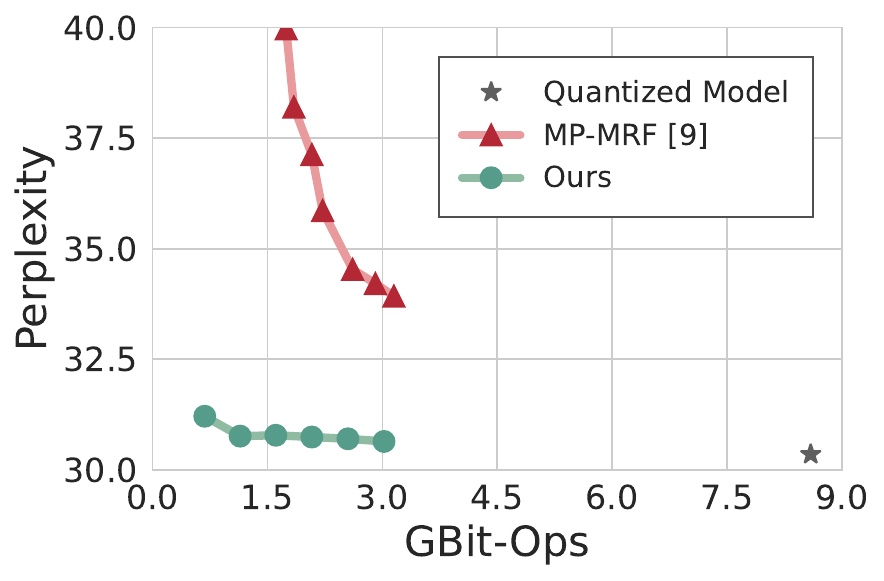}
        \caption{NLP task}
    \end{subfigure}
    \hfill
    \caption{Performance comparison between our work and prior work for (a) CV task and (b) NLP task.}
    \label{fig: comparison}
\end{figure}

We compare our work with MP-MRF proposed by Energon\cite{zhou2022energon}, adopting the range of $[-0.2, 0.2]$ with an interval of 0.1 for the search space of hyperparameters as the paper suggests. The MP-MRF method is implemented with the model quantized to 8-bit. The model performance and the number of bit operations are plotted in Fig. \ref{fig: comparison}. We calculate the number of bit operations by aggregating the number of MACs during the computation of $Q\cdot K^T$ and $P\cdot V$. The number of bit operations is then obtained by multiplying the bit-width of the multiplicand and the multiplier. 
% Take 8-bit multiplication, for example, the number of bit operations is $8\ bits\times 8\ bits=64\ bit$-$ops$.
\par
There are 2 key findings in Fig. \ref{fig: comparison}: 1) In the CV task, the performance of our work surpasses Energon with lower bit operations and accuracy degradation. In the marginal case, our work can bypass around 85.16\% keys and save 76.37\% bit operations with an accuracy drop of only 0.87\%. As for the NLP task, our work achieves promising results, with a mere 10.09\% of the keys attended leading to a perplexity increase of 0.86, reducing 92.06\% bit operations. 2) In the CV task, with some keys pruned, the accuracy even increases, which is also observed in Energon. As Energon argued, removing irrelevant keys helps the model be more concentrated on the remaining ones, which would be helpful in question-answering tasks. Our finding tells that this phenomenon also applies to image classification tasks. %(3) Our algorithm finds the solutions nearing the given target pruning ratio. While there is a slight discrepancy between the target pruning ratio and the found solution, we attribute this variance to the difference between the training and test sets. Nonetheless, our algorithm provides great control over the pruning ratio.

\subsection{Approximate Dot Product}
\begin{figure}[t]
    \centering
    \hfill
    \begin{subfigure}[b]{0.47\linewidth}
        \centering
        \includegraphics[width=\textwidth]{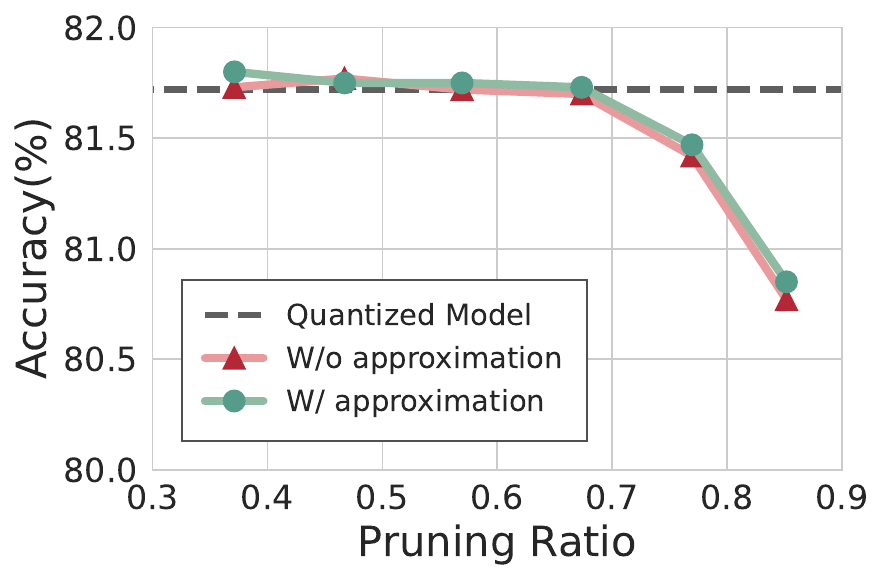}
        \caption{CV task}
    \end{subfigure}
    \hfill
    \begin{subfigure}[b]{0.47\linewidth}
        \centering
        \includegraphics[width=\textwidth]{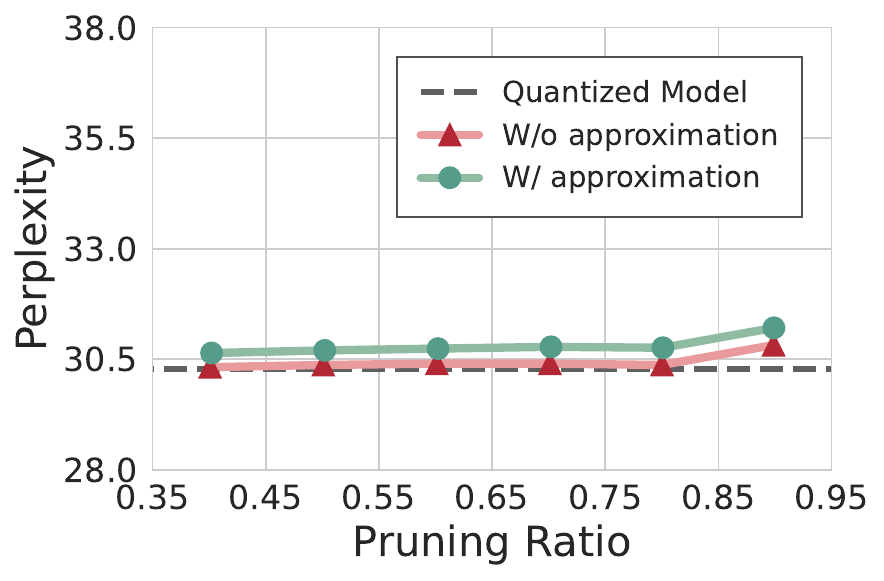}
        \caption{NLP task}
    \end{subfigure}
    \hfill
    \caption{Performance comparison before and after approximation for (a) CV task and (b) NLP task.}
    \label{fig: approx}
\end{figure}

As mentioned in subsection III-A-c, instead of calculating the complete dot product, we skip the negligible part of the multiplication between the LS4B and define the reduced computation as the approximate dot product. In this subsection, we conduct the experiment of the performance with and without the approximation. As Fig. \ref{fig: approx} shows, in both the CV and the NLP tasks, approximation has a minor influence on the model performance. Therefore, skipping the LS4B multiplication part can be seen as effective in reducing the computation without hurting the performance.

\begin{figure}[t]
    \centering
    \includegraphics[width=\linewidth]{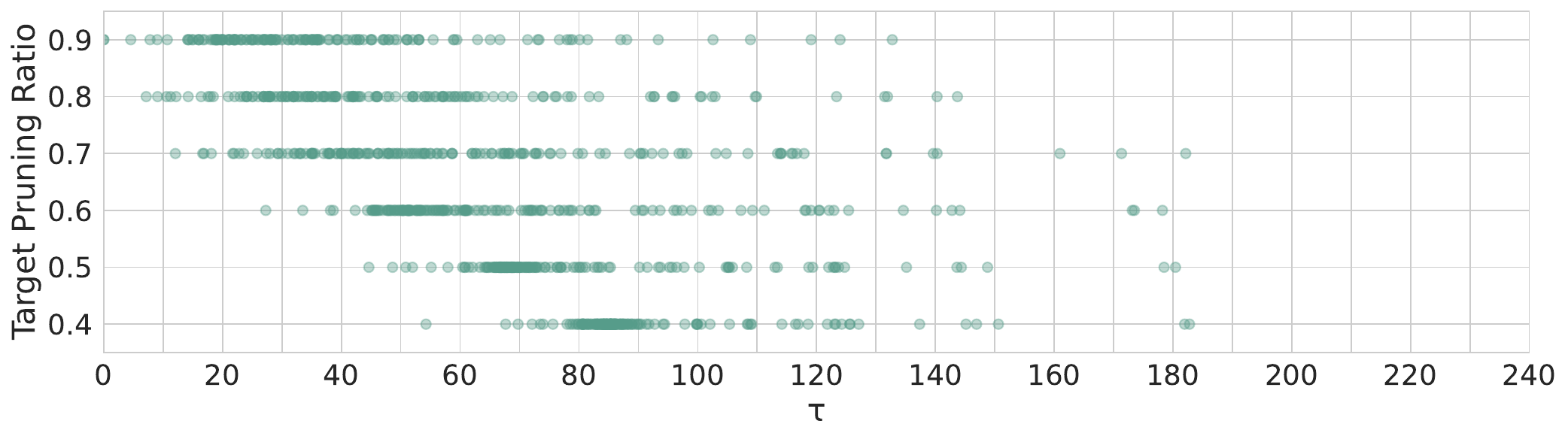}
    \caption{Visualization of trainable threshold under different target pruning ratios.}
    \label{fig: visualization}
\end{figure}

\subsection{The Necessity of Trainable Threshold}
To justify the necessity of the trainable threshold, we visualized the trained threshold for the CV task in Fig. \ref{fig: visualization}. To be more specific,  each head is paired with a hyperparameter $\tau$. In the DeiT-B16 model, with 12 blocks and 12 heads per block, there are 144 hyperparameters for the entire model. In Fig. \ref{fig: visualization}, for each target pruning ratio, we plot all 144 hyperparameters horizontally. In other words, there are 144 points in a horizontal plot. The plot gives the following observations: 1) As the target pruning ratio increases, the derived hyperparameter $\tau$ tends to decrease, aligning with our expectations that a smaller $\tau$ prunes more keys. 2) The hyperparameter $\tau$ varies across a broad range, making the exhaustive search impractical to identify the same configuration. Thus, under our proposed method, differential threshold, and head-wise hyperparameter tuning, the trainable threshold is essential for finding an optimal configuration.

\section{Conclusion}
In this paper, we propose \fulltitle\ (\abbrtitle). \abbrtitle\ takes advantage of the sparsity in Multi-Head Attention to reduce the computation by filtering the unimportant keys in advance. With the low-precision dot product and computation reuse mechanism, \abbrtitle\ minimizes the filter overhead. Also, the trainable threshold provides a systematic way for granular hyperparameter adjustment. Experimental results on both NLP and CV tasks show that \abbrtitle\ can filter up to 85.16\% keys with only a 0.87\% accuracy drop in the CV task and 89.91\% keys with a 0.86 perplexity increase in the NLP task.

\section*{Acknowledgments} This work was supported by the National Science and Technology Council of Taiwan under Grants of MOST 111-2218-E-002-018-MBK and NSTC 112-2218-E-002-025-MBK.

\bibliographystyle{IEEEtran}
\bibliography{refs}

\end{document}